\documentclass[journal]{IEEEtran}
\ifCLASSINFOpdf
\else
\fi

\usepackage{amsmath}
\usepackage{textcomp}
\usepackage{amsthm}
\usepackage{amssymb}
\usepackage{stfloats}
\usepackage{color}
\usepackage{engord}
\usepackage{graphicx}
\usepackage{mathtools}
\usepackage{multicol}
\usepackage[justification=centering]{caption}
\usepackage{autobreak}%公式自动断行
\allowdisplaybreaks[4]%跨页自动断页 括号中的参数可为1,2,3,4。数值越大,执行跨页的强度越大
\usepackage{caption}
\usepackage{graphicx}
\usepackage{float} 
\usepackage{subcaption}
\usepackage{stfloats}
\usepackage{cite}
\usepackage{lipsum} 
\usepackage{array}
\usepackage{autobreak}%公式自动断行
\allowdisplaybreaks[0]%跨页自动断页

\allowdisplaybreaks % 允许公式跨页显示

\begin{document}
% \title{\LARGE Statistical Guarantee Optimization for Supporting xURLLC\\ in ISAC-enabled V2I Networks with Short Packets}
%\title{\LARGE Supporting xURLLC in ISAC-enabled V2I Networks with Short Packets: Statistical AoI-Guaranteed Power Allocation}
% \title{\LARGE Supporting xURLLC in ISAC-enabled V2I Networks with Short Packets: A Stochastic Network Calculus Perspective}
% \title{\LARGE Statistical AoI Provisioning Analysis for Supporting xURLLC \\in ISAC-enabled V2I Networks}
%\title{\LARGE Enhancing xURLLC with Statistical AoI Guarantees \\ in ISAC-enabled V2I Networks}
\title{\LARGE Statistical AoI Guarantee Optimization for Supporting xURLLC\\in ISAC-enabled V2I Networks}

\author{Yanxi~Zhang, Mingwu~Yao, Qinghai~Yang, Dongqi~Yan, Xu~Zhang, Xu~Bao, and Muyu~Mei% <-this % stops a space

%\vspace{-0.3em} 
%投稿信息↓
\thanks{Yanxi Zhang, Mingwu Yao, Qinghai Yang, Dongqi Yan and Xu Zhang are with the State Key Laboratory of Integrated Services Networks, School of the Telecommunication Engineering, Xidian University, Xi’an 710071, China (e-mail: yanxi.zhang@stu.xidian.edu.cn; mwyao@xidian.edu.cn; qhyang@xidi\\an.edu.cn; 22011110194@stu.xidian.edu.cn; xuzhang265@stu.xidian.edu.cn). 

Xu Bao and Muyu Mei are with the School of Computer Science and Communication Engineering, Jiangsu University, Zhenjiang 212013, China (e-mail: xbao@ujs.edu.cn; mei\_muyu@163.com).} }

\maketitle


\begin{abstract}
This paper addresses the critical challenge of supporting next-generation ultra-reliable and low-latency communication (xURLLC) within integrated sensing and communication (ISAC)-enabled vehicle-to-infrastructure (V2I) networks. We incorporate channel evaluation and retransmission mechanisms for real-time reliability enhancement. Using stochastic network calculus (SNC), we establish a theoretical framework to derive upper bounds for the peak age of information violation probability (PAVP) via characterized sensing and communication moment generation functions (MGFs). By optimizing these bounds, we develop power allocation schemes that significantly reduce the statistical PAVP of sensory packets in such networks. Simulations validate our theoretical derivations and demonstrate the effectiveness of our proposed schemes.
\end{abstract}

\begin{IEEEkeywords}
Integrated sensing and communication, vehicle to infrastructure, next-generation ultra-reliable and low-latency communication, age of information, stochastic network calculus.
\end{IEEEkeywords}

% \fi
\IEEEpeerreviewmaketitle
%\vspace{-1em}
\section{Introduction}
 The rapid advancement of Vehicle-to-Everything (V2X) technology is revolutionizing modern transportation systems by enabling seamless communication between vehicles and various entities. V2X encompasses a broad spectrum of interactions, including Vehicle-to-Vehicle (V2V), Vehicle-to-Infrastructure (V2I), Vehicle-to-Network (V2N), and Vehicle-to-Pedestrian (V2P) \cite{ZhongEmpoweringV2XNetwork2022}. Among these, V2I plays a pivotal role in facilitating applications such as autonomous driving, efficient traffic management, and enhanced online services. Despite the promising benefits of V2I, the burgeoning market presents significant challenges. Traditional independently designed communication and sensing systems require a wide array of diverse resources, exacerbating resource congestion. To address this issue, ISAC emerges as a key technology in the development of 6G networks \cite{LiuIntegratedSensingCommunications2022}. ISAC integrates sensing and communication functionalities, leveraging shared hardware and resources to deliver both services efficiently. It offers substantial advantages for V2I networks by improving spectral efficiency and enabling hardware reuse. To fully capitalize on these advantages, effective power allocation is crucial as it balances the energy requirements between sensing and communication, thereby optimizing overall system performance. Recent studies have shown that innovative power allocation schemes in ISAC systems can enhance both communication and sensing capabilities while maintaining low power consumption \cite{Liu2023Performance,DuIntegratedSensingCommunications2023}. However, developing a unified theoretical framework for ISAC systems in V2I networks, along with basic performance limits and optimal schemes, remains a significant challenge.

V2I networks depend heavily on accurate and timely provision of high-quality environmental information. Given the highly dynamic nature of vehicular environments, the validity period of such information is inherently short, making it extremely sensitive to real-time performance. To meet these stringent requirements, xURLLC in 6G is evolving toward a broader set of real-time metrics, advancing beyond the latency-centric metrics prevalent in 5G \cite{Lopez2023Statistical}. A key metric gaining prominence within xURLLC is the AoI, which emphasizes data freshness. Achieving high real-time performance typically involves using short packets, where Shannon capacity is not applicable. Instead, finite blocklength coding (FBC) theory has been developed to approximate the maximum achievable data rate while accounting for decoding error probabilities (DEP) associated with short packets \cite{Lopez2023Statistical}. Furthermore, ensuring reliable transmission is paramount for road safety in V2I networks; however, achieving this reliability with short packets remains challenging due to their limited error tolerance. Despite its critical importance, there is a notable scarcity of literature exploring ISAC-enabled V2I networks with short packets to support xURLLC. The few studies that may be relevant tended to focus on latency \cite{Zhao2024Joint,Ding2022Joint}. However, a statistical understanding of AoI for supporting xURLLC within ISAC-enabled V2I networks is still lacking.

Several studies have been conducted on the statistical analysis of AoI. However, most of them rely on simplistic assumptions, typically assuming the packet transmission times to be constant \cite{Zhang2021AoI} or follow fixed distributions \cite{Champati2021Statistical}, which do not reflect the time-varying dynamics of V2I networks. A recent study \cite{Zhong2023Stochastic} considers time-varying channel in the internet of everything (IoE), assuming the channel rate remains constant during each packet transmission. While this represents a step forward, it falls short in certain situations. Specifically, when the time-varying channel rate is excessively low, the theoretical packet transmission time becomes excessively large and exceeds the coherence time, thus invalidating the assumption and leading to inaccurate performance analysis. To address this, our work introduces a channel evaluation mechanism that filters out unacceptable rates in time-varying channels. This not only strengthens the robustness of the theoretical analysis, but also enhances the reliability of V2I networks, demonstrating both theoretical and practical significance.

In this work, we tackle the issue of minimizing the statistical PAVP of sensory packets in a power-limited ISAC-enabled V2I network while ensuring reliability. We present an analysis framework for this network that integrates channel evaluation and retransmission mechanisms to improve AoI performance and reliability. First, we employ SNC to establish a theoretical upper bound framework based on MGFs. Then, we characterize the sensing and communication capabilities under varying channel conditions. By integrating these characterized MGFs into our proposed framework, we develop a power allocation strategy that optimizes the statistical PAVP of sensory packets.

\begin{figure}
    \centering{
    \includegraphics[width=7cm]{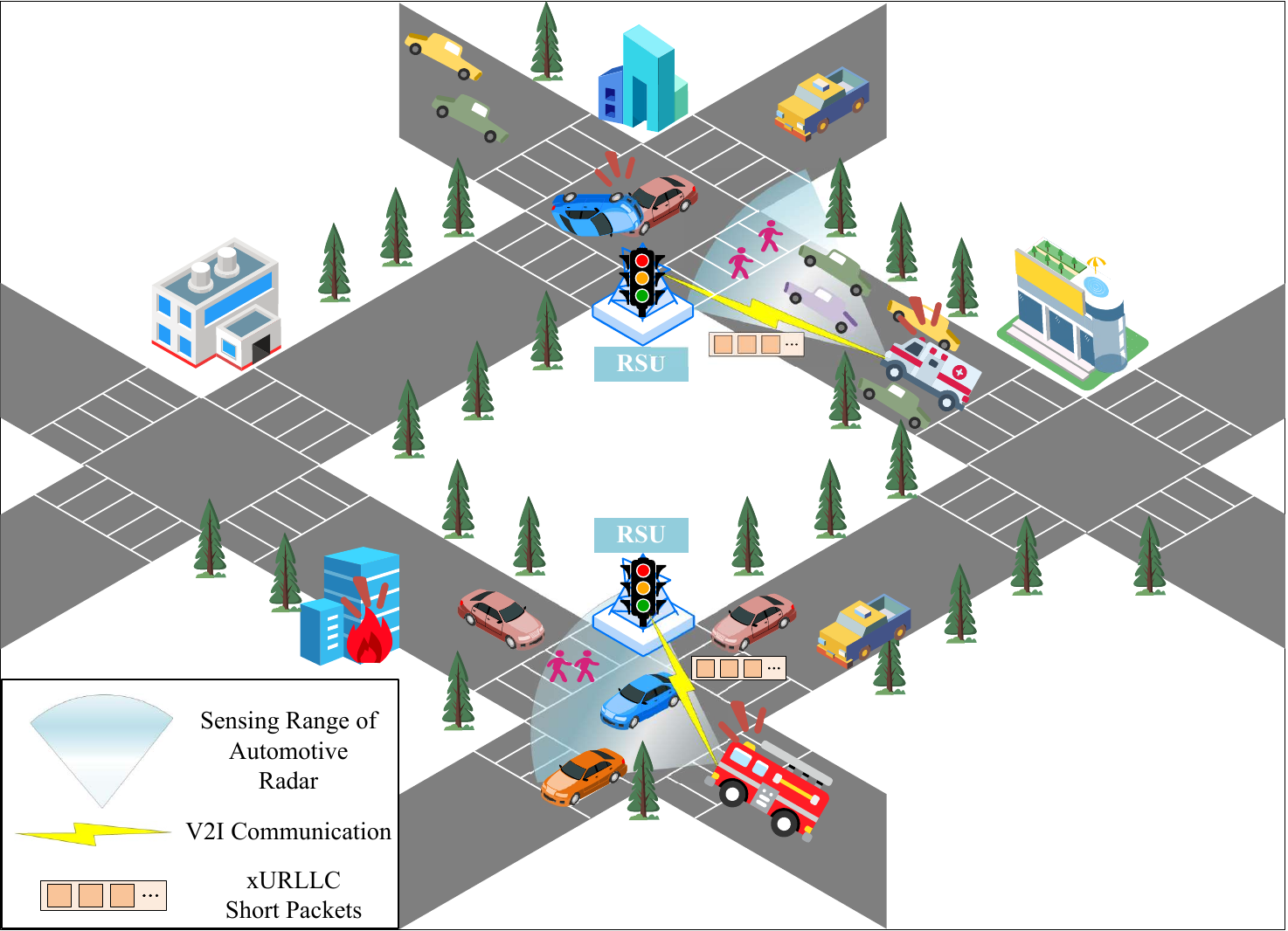}
    \captionsetup{justification=raggedright,singlelinecheck=false} % 设置标题左对齐
    \caption*{\small Fig. 1: The ISAC-enabled V2I networks.}
    \label{fig:1}}
\vspace{-1.5em}
\end{figure}

\vspace{-0.4em}
\section{System Model}
\vspace{-0.1em}

\subsection{Network Model}
As illustrated in Fig. 1, we consider an emergency response scenario that leverages an ISAC-enabled V2I network to improve the efficiency and safety of emergency vehicles (EVs), such as ambulances and fire trucks. Each EV is equipped with a front-mounted long-range automotive radar for road condition detection. The ISAC-enabled system on the EV allocates orthogonal frequency resources to sensing and communication, enabling simultaneous operations without mutual interference. The system conducts sensing detection, aggregates critical data into short packets, then transmits them wirelessly to a nearby roadside unit (RSU). Upon receipt, the RSU can dynamically alter traffic signals and instruct other vehicles to slow down or reroute, thereby facilitating a swift passage for the EV through congested areas. This system operates in three distinct phases:

% It first conducts sensing detection, aggregates the critical sensory data into short packets, then transmits these packets wirelessly to the road side unit (RSU). The received packets enable the RSU to dynamically alter traffic signals and instruct other vehicles to slow down or reroute, facilitating a swift passage for the EV through congested areas. The operation cycle of such system is divided into three phases: 

\noindent\textit{1) Phase I (Radar Scanning and Packet Generation)}

When an EV is dispatched on a mission, its automotive radar performs periodic scans at intervals of $T$ to detect road conditions. Each successful detection generates a short sensory packet containing critical information with xURLLC requirements. These short packets are of fixed length with $L$ bits, indexed by $n \in \mathbb{N}$. Upon generation, these packets are stored in the EV's buffer, awaiting transmission.

% Through effective management of the packets in the buffer, the system can adjust transmission strategies based on channel conditions, thereby minimizing unnecessary communication delays.

\noindent\textit{2) Phase II (\textit{Channel State Evaluation})}

In ISAC-enabled V2I networks, unreliable information can lead to severe consequences, such as traffic accidents. Thus, ensuring transmission reliability is critical. We assume that the perfect channel state information (CSI) is available at the RSU \cite{Li2023Effective}. Before transmission, the system evaluates the instantaneous channel signal-to-noise ratio (SNR). If the current SNR exceeds a threshold $\tau$, the channel is deemed acceptable for transmission, and the packet transmission process in \textit{Phase III} proceeds. Otherwise, the transmission is deferred by an interval $\varpi$, after which the SNR is re-evaluated. This process continues until an acceptable channel state is reached, significantly improving the reliability of transmission.

\noindent\textit{3) Phase III (Transmission and Retransmission under FBC)}

Short sensory packets are encoded in the FBC regime and transmitted over a wireless Rayleigh fading channel to the RSU under the first-come-first-serve (FCFS) discipline. It is assumed that each packet can be transmitted by the EV within the timescale of small-scale channel fading. In this regard, the channel gain between the EV and the RSU is denoted as $h$, where $h$ is independently and identically distributed (i.i.d.) across different packets. Based on FBC theory \cite{Lopez2023Statistical}, the achievable data rate can be approximated as
\begin{align}
R\left(\gamma(h)\right) \approx \frac{W}{\ln 2}\left[\ln \left(1+\gamma(h)\right)\!-\!\sqrt{\frac{V\left(\gamma(h)\right)}{N}} Q^{-1}\left(\epsilon\right)\right], 
\label{a}
\end{align}
where $\gamma(h)$ is the SNR, $\epsilon$ is the DEP, $N$ is the blocklength, $Q(x)$ is the Gaussian Q-function, and the channel dispersion is $V\left(\gamma(h)\right)=1-{\left(1+\gamma(h)\right)^{-2}}.$ The RSU notifies the EV of transmission success or failure via immediate ACK/NACK feedback. Any packet in error is discarded, and the EV attempts retransmission after the next several intervals where the channel state is acceptable. This process repeats until successful reception, enhancing transmission reliability. 

% The RSU indicates transmission success or failure to the EV by sending an ACK or NACK, which are assumed to be returned instantaneously, allowing the EV to accurately track each packet's reception status.

\vspace{-0.7em}
\subsection{Performance Metric and Problem Statement}
\vspace{-0.1em}
Since the information in the sensory packets is crucial to road safety, it should not only meet the stringent xURLLC requirements, but also adequately reflect the current environment to allow immediate decision-making. The AoI at the RSU receiver serves as a key metric to evaluate this timeliness, capturing both the freshness of sensing and transmission. Specifically, we focus on the peak AoI (PAoI), which provides insights into the worst-case scenario for data freshness. The PAVP of the sensory packet $n$ is defined as $\mathbb{P}\left[\delta_A(n)>\zeta\right]$, where $\delta_A(n)$ denotes the PAoI of sensory packet $n$, and $\zeta$ is a threshold for the maximum tolerable PAoI by the receiver.

Consider the ISAC-enabled V2I network under a power constraint $P_t$:
$P_t=P_s + P_c,$
where $P_s$ is the power for sensing and $P_c$ is the power for communication. Our objective is to minimize the statistical PAVP of sensory packets in this network given the total power constraint, leading to the formulation of problem $\mathcal{P}$ :
\begin{align}
\mathcal{P}: \min _{0 \leq \alpha \leq 1}  \mathbb{P}\left[\delta_A(n)>\zeta\right],
\label{z}
\end{align}
where $\alpha$ is the power allocation coefficient, denoting the ratio of communication power to the total power: $\alpha=P_c/P_t$.

\vspace{-0.5em}
\section{The Analysis Framework for PAVP}
In the previous section, we have identified the problem $\mathcal{P}$ that we aim to solve. However, directly computing $\mathbb{P}\left[\delta_A(n)>\zeta\right]$ is challenging due to the involvement of dynamic stochastic processes in sensing and communication, as well as complex queue evolution. Therefore, we propose applying SNC to derive an upper bound on the PAVP of sensory packets in the ISAC-enabled V2I network, providing a theoretical framework to investigate the effects of stochastic parameters on network performance.

We denote $T^A(n)$ and $T^D(n)$ as the arrival and departure times of the packet $n$, respectively. For all $n \geq 1$, causality implies that $T^D(n) \geq T^A(n)$. Without loss of generality, we initialize with $T^A(0)=T^D(0)=0$. We denote $T^A(v, n)$ as the inter-arrival interval and $T^D(v, n)$ as the departure time interval between the $v^{\text{th}}$ and the $n^{\text{th}}$ packet. We have $T^A(v, n)=T^A(n)-T^A(v)$ and $T^D(v, n)=T^D(n)-T^D(v)$.

A common definition of AoI at time $t$ is $\Delta(t)=t-$ $\max _{n \geq 1}\{T^A(n): T^D(n)<t\}$. Thus, the PAoI of sensory packet $n \geq 1$ is given by
\begin{align}
    \delta_A(n)&=T^D(n+1)-T^A(n) \nonumber\\
    &=T^D(n+1)-T^A(n+1)+T^A(n+1)-T^A(n) \nonumber\\
    &=T^D(n+1)-T^A(n+1)+T^A(n,n+1).
    \label{c}
\end{align}

Given that PAoI is determined by the timestamps of received sensory packets at the destination, we model the ISAC-enabled V2I network using packet-based SNC theory. In the following, we present some fundamental representations and results.

\textit{Definition 1:} A system with the arrival process $T^A(n)$ and the departure process $T^D(n)$ is considered an exact $T^S(v, n)$ server if for all $n \geq 1$:
$$
T^D(n)=\max _{1 \leq v \leq n}\{T^A(v)+T^S(v, n)\}.
$$

\textit{Lemma 1:} For an FCFS system, the relationship between $T^D(n), T^A(n)$ and $T^S(n)$ is
\begin{equation}
T^D(n)=\max _{1 \leq v \leq n}\left\{T^A(v)+\sum_{\ell=v}^n T^S(\ell)\right\},
\label{d}
\end{equation}
which is an exact $T^S(v, n)=\sum_{\ell=v}^n T^S(\ell)$ server.

\textit{Proof:}
Let $Z(n)$ denote the time when packet $n$ starts service, and $L(n)$ denote its service time. For packet 1, we have $Z(1)=T^A(1)$ since it starts service upon arrival. Under FCFS policy, any subsequent packet $n \geq 2$ starts service at
\begin{equation}
Z(n)=\max \{T^A(n), Z(n-1)+T^S(n-1)\},
\label{f}
\end{equation}
indicating that packet $n$ starts service either upon arrival or after the previous packet $n-1$ has completed its service.

By recursive insertion of Eq. \eqref{f}, we have
\begin{equation}
Z(n)=\max _{1 \leq v \leq n}\left\{T^A(v)+\sum_{\ell=v}^{n-1} T^S(\ell)\right\}.
\label{g}
\end{equation}

Since $T^D(n)=Z(n)+L(n)$, substituting Eq. \eqref{g} yields 
\begin{equation}
T^D(n)=\max _{1 \leq v \leq n}\left\{T^A(v)+\sum_{\ell=v}^n T^S(\ell)\right\}.
\label{h}
\end{equation}

Hence, the system operates as an exact $T^S(v, n)=\sum_{\ell=v}^n T^S(\ell)$ server, completing the proof of \textit{Lemma 1}. \hspace*{\fill}\qed

By inserting Definition 1 into Eq. \eqref{c}, it follows readily that
$$
\delta_A(n) \leq \max _{1 \leq v \leq n+1}\{T^S(v, n+1)-T^A(v, n+1)\} +T^A(n,n+1).
$$

Since the PAVP cannot be computed directly, we envelope it by using the MGF. The MGF of a non-negative random variable $X$ is defined as $\mathbf{M}_X(\theta)=\mathbb{E}\left[e^{\theta X}\right]$, where $\theta>0$ is a free parameter. With the aid of MGF, an upper bound on PAVP is presented in the following lemma, which holds under the stability condition $\mathbf{M}_{\mathcal{T}^A(n, n+1)}(-\theta) \mathbf{M}_{\mathcal{T}^{\mathcal{S}}(n+1)}(\theta)< 1$.

% \textbf{Theorem 1}: Given the peak AoI threshold $A_{\mathrm{th}}$, the upper bound on the peak AoI violation probability $p^{(\mu, \mathrm{Aol})}$ for our proposed AoI-driven and URLLC-enabled schemes in the finite blocklength regime is given as follows:
% $$
% p^{(\mu, \mathrm{Aol})} \leq e^{-\theta A d} \mathcal{K}(\theta, \mu)
% $$
% where
% $$
% \begin{aligned}
% \mathcal{K}(\theta, \mu) \triangleq \mathbf{M}_{\mathcal{T}(\mu, \mu+1)}(\theta) & {\left[\sum_{v=1}^{\mu+1} \mathbf{M}_{\mathcal{T}^{\mathrm{s}}(v, \mu+1)}(\theta)\times \mathbf{M}_{\mathcal{T}^A(v, \mu+1)}(-\theta)\right]} .
% \end{aligned}
% $$

\textit{Lemma 2:} If $\theta>0$, for a given threshold $\zeta$, the PAVP upper bound is given as
\begin{align}
  \mathbb{P}\left[\delta_A(n)>\zeta\right] \leq \frac{e^{-\theta \zeta} \mathbf{M}_{\mathcal{T}^A(n, n+1)}(\theta)}{[\mathbf{M}_{\mathcal{T}^{\mathrm{S}}(n)}(\theta)]^{-1}-\mathbf{M}_{\mathcal{T}^A(n, n+1)}(-\theta)}.
  \label{i}
\end{align}

% \textbf{Corollary 1.} If $\theta>0$, for a given PAVP $\varrho$, the PAoI upper bound is given as follows:
% \begin{align}
%   \zeta = -\frac{1}{\theta}\ln{\left[\frac{\varrho([\mathbf{M}_{\mathcal{T}^{\mathrm{S}}(n)}(\theta)]^{-1}-\mathbf{M}_{\mathcal{T}^A(n, n+1)}(-\theta))}{\mathbf{M}_{\mathcal{T}^A(n, n+1)}(\theta)}\right]}. \nonumber
% \end{align}

\textit{Proof:} The MGF of the PAoI $\mathbf{M}_{\delta_A(n)}(\theta)$ can be derived as
\begin{align}
&\mathbf{M}_{\delta_A(n)}(\theta)=\mathbb{E}\left[e^{\delta_A(n)\theta}\right] \nonumber\\
\leq &\sum_{v=1}^{n+1} \mathbb{E}\left[e^{\theta T^S(v, n+1)}\right] \mathbb{E}\left[e^{-\theta T^A(v, n+1)}\right]\mathbb{E}\left[e^{\theta T^A(n, n+1)}\right] \nonumber,
\end{align}
where in the last step, we use Boole's inequality.

Assuming that the inter-arrival times $T^{A}(n, n+1)$ are i.i.d. for each sensory packet, we have
\begin{align}
\mathbb{E}\left[e^{-\theta T^A(v, n+1)}\right]&=\mathbb{E}\left[\prod_{\ell=v}^{n} e^{-\theta T^{A}(\ell, \ell+1)}\right] \nonumber \\
&=\left(\mathbb{E}\left[e^{-\theta T^{A}(n, n+1)}\right]\right)^{(n-v+1)}.
\label{j}
\end{align}

Similarly, assuming that the service times $T^{\mathrm{S}}(n)$ are i.i.d. for each sensory packet, we can derive the MGF of the cumulative service time as
\fontsize{9.5pt}{11.5pt}\selectfont
\begin{align}
\mathbb{E}\!\left[e^{\theta T^S(v, n+1)}\!\right]\!\!=\!\mathbb{E}\!\left[\prod_{\ell=v}^{n+1} \! e^{\theta T^{\mathrm{s}}(\ell)}\!\right]\!\! =\!\left(\mathbb{E}\!\left[e^{\theta T^{\mathrm{s}}(n)}\right]\right)^{\!(n-v+2)}.
\label{v}
\end{align}
\normalsize

Using Eqs. \eqref{j} and \eqref{v}, we can rewrite $\mathbf{M}_{\delta_A(n)}(\theta)$ as
\begin{align}
&\mathbf{M}_{\delta_A(n)}(\theta) \nonumber\\
\leq&\sum_{v=1}^{n+1} \mathbb{E}\left[e^{\theta T^S(v, n+1)}\right] \mathbb{E}\left[e^{-\theta T^A(v, n+1)}\right]\mathbb{E}\left[e^{\theta T^A(n, n+1)}\right] \nonumber\\
=&\mathbb{E}\left[e^{\theta T^A(n, n+1)}\right]\mathbb{E}\left[e^{\theta T^{\mathrm{s}}(n)}\right] \nonumber\\
&\quad \quad \quad \quad \quad \cdot \sum_{v=1}^{n+1}\left(\mathbb{E}\left[e^{\theta T^{\mathrm{s}}(n)}\right]\mathbb{E}\left[e^{-\theta T^{A}(n, n+1)}\right]\right)^{(n-v+1)} \nonumber \\
\leq &\mathbb{E}\!\left[e^{\theta T^A(n, n\!+\!1)}\!\right]\!\mathbb{E}\!\left[e^{\theta T^{\mathrm{s}}(n)}\!\right] \!\sum_{v=0}^{\infty}\!\left(\!\mathbb{E}\!\left[e^{\theta T^{\mathrm{s}}(n)}\!\right]\!\mathbb{E}\!\left[e^{-\theta T^{A}(n, n\!+\!1)}\!\right]\!\right)^{\!v} \nonumber \\
=&\frac{\mathbb{E}\left[e^{\theta T^{A}(n, n+1)}\right] \mathbb{E}\left[e^{\theta T^{\mathrm{S}}(n)}\right]}{1-\mathbb{E}\left[e^{-\theta T^{A}(n, n+1)}\right] \mathbb{E}\left[e^{\theta T^{\mathrm{S}}(n)}\right]},
\label{k}
\end{align}
where in the last step, we utilize the stability condition:
$
\begin{aligned}
\mathbb{E}\left[e^{-\theta T^{A}(n, n+1)}\right] \mathbb{E}\left[e^{\theta T^{\mathrm{S}}(n)}\right]<1 .
\end{aligned}
$

With Chernoff's inequality $\mathbb{P}[X \geq x]\leq e^{-\theta x}\mathbb{E}[e^{\theta X}]$, we can derive the PAVP upper bound as
\begin{align}
\mathbb{P}\left[\delta_A(n)>\zeta\right]
&\leq e^{-\theta \zeta} \mathbf{M}_{\mathcal{P}^{\text {Aol }}(n)}(\theta) \nonumber\\
&=\frac{e^{-\theta \zeta} \mathbb{E}\left[e^{\theta T^{A}(n, n+1)}\right] \mathbb{E}\left[e^{\theta T^{\mathrm{S}}(n)}\right]}{1-\mathbb{E}\left[e^{-\theta T^{A}(n, n+1)}\right] \mathbb{E}\left[e^{\theta T^{\mathrm{S}}(n)}\right]} \nonumber\\
&=\frac{e^{-\theta \zeta} \mathbb{E}\left[e^{\theta T^{A}(n, n+1)}\right]}{\left(\mathbb{E}\left[e^{\theta T^{\mathrm{S}}(n)}\right]\right)^{-1}-\mathbb{E}\left[e^{-\theta T^{A}(n, n+1)}\right]}\nonumber\\
& = \frac{e^{-\theta \zeta} \mathbf{M}_{\mathcal{T}^A(n, n+1)}(\theta)}{[\mathbf{M}_{\mathcal{T}^{\mathrm{S}}(n)}(\theta)]^{-1}-\mathbf{M}_{\mathcal{T}^A(n, n+1)}(-\theta)},
\label{l}
\end{align}
which completes the proof of \textit{Lemma 2}. \hspace*{\fill}\qed

% Subsequently, Corollary 1 is obtained by equating the upper bound in Lemma 3 with a given PAVP $\varrho$ and solving for $\zeta$.

Using SNC, we have derived an upper bound on the PAVP, providing a robust theoretical framework for network performance analysis. Building upon this theoretical foundation, the next section will provide a detailed MGF characterization for the ISAC-enabled V2I network.

%\vspace{-0.5em}
\section{Joint Analysis and Optimization for PAVP of Sensory Packets in the ISAC-enabled V2I Network}
As presented in \textit{Lemma 2}, the upper bound on PAVP involves computing the MGF for the inter-arrival time and service time of the packets. This section characterizes MGFs for sensing and communication in the ISAC-enabled V2I network, followed by a joint power allocation optimization to achieve optimal PAVP.

%\vspace{-0.5em}
\subsection{MGF Characterization for Packet Inter-arrival Time}
This subsection characterizes the MGF of the packet service time in the ISAC-enabled V2I network. As stated in Sec. II, each successful radar detection generates a sensory packet containing critical information. The SDP is defined as the probability that the detection is successful only if the received sensing SNR $\gamma_s$ at the vehicle exceeds a sensing threshold $d$. This metric of radar performance is given by
\begin{equation}
\mathbb{P}_s(d)=\mathbb{P}[\gamma_s>d]=\mathbb{P}\left[\frac{P_{e}}{N_s}>d\right],
\label{w}
\end{equation}
where $P_{e}$ is the echo power and $N_s=W \varsigma \chi \varphi$ is the sensing noise power. Denote $\varsigma=1.38 \times 10^{-23}$ w/s as the Boltzmann's constant, $\chi=290 \mathrm{~K}$ as the standard temperature, $W$ as the equivalent noise bandwidth and $\varphi=10$ dB as the system loss factor. 
According to \cite{Series2018Systems}, $P_{e}$ is expressed as
\begin{equation}
P_{e}=\frac{P_s \Gamma_t \Gamma_r \sigma^2 \rho}{(4 \pi)^3 D^{2 \kappa}},
\label{m}
\end{equation}
where $\Gamma_t$ and $\Gamma_r$ are the transmitting and receiving antenna gains respectively, $\sigma$ is the radar wavelength, $D$ is the maximum ranging distance, and $\kappa$ is the path-loss exponent. The radar cross section (RCS) is denoted by a random variable $\rho$.

The radar echo from a moving target typically fluctuates from scan to scan, which is crucial for the precision of radar performance. The Swerling I model effectively evaluates this fluctuation by assuming that radar echoes are constant during a single scan but vary independently between scans. The probability density function (PDF) for $\rho$ in this model is

\begin{equation}
f(\rho)=\frac{1}{\bar{\rho}} e^{-\frac{\rho}{\bar{\rho}}},\ \rho \geq 0,
\label{n}
\end{equation}
where $\bar{\rho}$ is the average of RCS. We now present the following lemma regarding the SDP with the Swerling I model.

\textit{Lemma 3:} For a given threshold $d$, the SDP of the EV with Swerling I model, denoted as $\mathbb{P}_s(\rho)$, is given by
\begin{equation}
\mathbb{P}_s(d)=\exp \left(-\frac{d(4 \pi)^3 D^{2 \kappa} N_s}{P_s \Gamma_t \Gamma_r \sigma^2 \bar{\rho}}\right).
\label{o}
\end{equation}

\textit{Proof:}
Insert Eq. \eqref{m} into Eq. \eqref{w}, $\mathbb{P}_s(d)$ is rewritten as
\begin{align}
\mathbb{P}_s(d)\!=\!\mathbb{P}\!\left[\frac{\frac{P_s \Gamma_t \Gamma_r \sigma^2 \rho}{(4 \pi)^3 D^{2 \kappa}}}{N_s}\!>d\right]\!\!=\!\mathbb{P}\!\left[\rho >\!\frac{d(4 \pi)^3 D^{2 \kappa}}{P_s \Gamma_t \Gamma_r \sigma^2}N_s\right]\!. 
\label{x}
\end{align}

Further, Eq. \eqref{x} is expressed as
$$
\mathbb{P}_s(d)=\mathbb{P}\left[\rho>\frac{d(4 \pi)^3 D^{2 \kappa}N_s}{P_s \Gamma_t \Gamma_r \sigma^2}\right].
$$
By using Eq. \eqref{n}, we obtain the result of \textit{Lemma 3}. \qed

In each radar scanning cycle $T$, the packet generation process can be described probabilistically. Specifically, there is a probability $\mathbb{P}_s(d)$ that a sensory packet is generated due to a successful radar scan, and a complementary probability $1\!-\!\mathbb{P}_s(d)$ that no sensory packet is generated. The packet generation process $\Lambda(d)$ of each scan can be modeled as
\begin{equation}
\Lambda(d)=\left\{\begin{array}{cl}
1, & \mathbb{P}_s(d), \\
0, &1-\mathbb{P}_s(d),
\end{array}\right.
\label{p}
\end{equation}
which represents a Bernoulli process with probability $\mathbb{P}_s(d)$. This leads to geometrically distributed inter-arrival times with parameter $\mathbb{P}_s(d)$. Using the MGF of a geometric distribution, the MGF for the packet inter-arrival time is given by
\begin{align}
\mathbf{M}_{\mathcal{T}^A(n, n+1)}(\theta)=\left(\left(e^{-\theta T}\!\!-1\right){\mathbb{P}_s(d)}^{-1}+1\right)^{-1}.
\label{q}
\end{align}

% \begin{align}
% \mathbf{M}_{\mathcal{T}^A(n, n+1)}(\theta)=\frac{\mathbb{P}_s(d)}{e^{-\theta T}-1+\mathbb{P}_s(d)},
% \label{q}
% \end{align}
%\vspace{-0.7em}
\subsection{MGF Characterization for Packet Service Time}
% Blocklength $N$ is an essential parameter of finite blocklength coding theory, which represents the number of symbols transmitted in one frame,
% $$
% N=T^d W
% $$
% where $T^d$ represents the data transmission time. $W$ is the bandwidth of each subcarrier. 

% In order to better approximate $V$, the value of $V$ is not approximated to the fixed value 1 as other works. Although this method increases the calculation difficulty, we could solve this issue by the RL method.
% We consider a discrete-time system in which the transmission time is divided into multiple slots. In each time slot, the transmission of control signal and data is required. We use $T^s$ to represent the TTI size, which is determined according to the service characteristics and channel conditions [29]. $T^c$ and $T^d$ respectively represent the control signal transmission time and data transmission time in one TTI. Then,
% $$
% T^s=T^c+T^d
% $$

% Different from other works [26], the $T^c$ in our work is not fixed but determined by the control overhead size and the transmission rate. The size of control signaling is directly related to the real-time SNR [29].
In this subsection, we characterize the MGF of the packet service time within the ISAC-enabled V2I network. As previously described in Sec. II, once a sensory packet is generated and stored in the EV's buffer, it goes through two phases before being successfully received by the RSU.

When a packet is ready to leave the buffer after queuing, the wireless channel state is evaluated. The decision to transmit or defer depends on whether the channel is in an acceptable state. Given that we are considering a Rayleigh fading channel and that the SNR is i.i.d. with an exponential distribution across packets, the probability $p$ that the wireless channel is in an acceptable state at each evaluation is given by $p=$ $\exp (-\tau N_c/P_c)$, where $N_c$ is the communication noise power.

Suppose a packet experiences $c_1$ deferrals (each of $ \varpi $ duration) before its first transmission attempt. If this transmission fails due to decoding errors and the packet is discarded, the EV waits for another $ c_2 $ intervals before attempting retransmission. This process repeats until the packet is successfully received. Let $ b $ be the number of failed transmission attempts before success on the $ (b+1)^{\text{th}}$ attempt, and let $ c = c_1 + c_2 + \cdots + c_{b+1} $ be the total number of intervals deferred across all attempts. The probability of a packet having a service time $S\left(\gamma(h)\right)$ in the ISAC-enabled V2I network for $b\geq 0, c \geq 0 $ is given by 
\begin{align}
&\mathbb{P}\left[ S\left(\gamma(h)\right)=c \varpi + (b + 1) \cdot \phi\left(\gamma(h)\right) \right] \nonumber\\
=&\binom{c + b}{c} \left(1 - p\right)^c  p^{b+1} \left(1 - \eta\right)^b  \eta,
\label{r}
\end{align}
where $ \eta=(1-\epsilon)^L$ is the probability of successful packet decoding, and $\phi\left(\gamma(h)\right)=L/R\left(\gamma(h)\right)$
is the time required for each transmission attempt of a packet in the FBC regime. 

Then, the MGF of the packet service time $\mathbf{M}_{T^{S}(n)}(\theta)$ can be derived as
\begin{align}
& \mathbf{M}_{T^{S}(n)}(\theta) \nonumber= \mathbb{E}\!\left[e^{\theta T^{S}}\right]\nonumber\\
% = &\mathbb{E}\left[\mathbb{E}\left[e^{\theta T^{S}} \mid h \geq \tau\right]\right]\nonumber\\
=& \mathbb{E}\!\left[\sum_{b=0}^\infty \! \sum_{c=0}^\infty \!\binom{c \!+\! b}{c}  e^{\theta\left(c\varpi+(b\!+\!1)\phi(\gamma(h))\right)} \!\left(1\!-\!p\right)^c \! p^{b+1} \!\left(1\!-\!\eta \right)^b \!\eta \right] \nonumber\\
=& \mathbb{E}\!\left[\sum_{b=0}^\infty e^{\theta(b+1)\phi(\gamma(h))} p^{b+1} \!\left(1\!-\!\eta \right)^b \!\eta \! \sum_{c=0}^\infty \! \binom{c \!+\! b}{c}  e^{\theta c\varpi} \!\left(1\!-\!p\right)^c \! \right] \nonumber\\
\overset{(a)}{=}& \mathbb{E}\!\left[\sum_{b=0}^\infty \frac{e^{\theta(b+1)\phi(\gamma(h))}p^{b+1} \left(1-\eta \right)^b \eta}{\left(1-e^{\varpi\theta}(1\!-\!p)\right)^{b+1}} \right] \nonumber\\
\overset{}{=}& \mathbb{E}\!\left[\frac{e^{\theta \phi(\gamma(h))}p \eta}{1-e^{\varpi\theta}(1-p)}\sum_{b=0}^\infty \left(\frac{e^{\theta \phi(\gamma(h))}p \left(1-\eta \right) \eta}{1-e^{\varpi\theta}(1-p)}\right)^b \right] \nonumber\\
=& \mathbb{E}\!\left[\frac{e^{\theta \phi(\gamma(h))}p \eta}{1-e^{\varpi\theta}(1-p)} \cdot \frac{1}{1- \frac{e^{\theta \phi(\gamma(h))}p \left(1-\eta \right) \eta}{1-e^{\varpi\theta}(1-p)} }\right] \nonumber\\
= & \mathbb{E}\!\left[\frac{ p \eta}{(1-e^{\varpi\theta}(1-p))e^{-\phi(\gamma(h))\theta}-p(1-\eta)}\right]\nonumber\\
\overset{(b)}{=} & \displaystyle\int_{\frac{\tau N_c}{P_c}}^{+\infty} \frac{p \eta e^{-h}}{(1-e^{\varpi\theta}(1-p))e^{-\phi(\gamma(h))\theta}-p(1-\eta)} \, \mathrm{d}h,
\label{s}
\end{align}
where in step (a), we simplify the inner sum using the Taylor series expansion; in step (b), we note that since the unacceptable states that less than $\tau$ have already been filtered out in prior \textit{Phase II} evaluations, the channel gain $h$ here follows a truncated exponential distribution.

% \textbf{Shannon Capacity}

% When the channel small-scale fading follows the Rayleigh distribution, the PDF of transmission time for an information update can be obtained as:
% $$
% \begin{aligned}
% f_{T^S(n)}(x) & =f_{|h|^2}\left(\frac{\left(2^{\frac{\lambda}{W}}-1\right) N_0 W}{P}\right) \cdot\left|\left[\frac{\left(2^{\frac{\lambda}{x}}-1\right) N_0 W}{P}\right]^{\prime}\right|  \\
% & =\frac{\ln 2 \cdot N_0 \lambda}{P} \cdot \frac{2^{\frac{\lambda}{x W}}}{x^2} \cdot \exp \left(\frac{\left(1-2^{\frac{\lambda}{x W}}\right) N_0 W}{P}\right)
% \end{aligned}
% $$
% Then, 
% \begin{align}
% & \mathcal{M}_{\psi}(1+\theta) \nonumber\\
% = & \displaystyle\int_0^{+\infty}\frac{\ln 2 \cdot N_0 \lambda}{P} \cdot \frac{2^{\frac{\lambda}{x W}}}{x^2} \cdot \exp \left(\frac{\left(1-2^{\frac{\lambda}{x W}}\right) N_0 W}{P}+\theta x\right) d x
% \end{align}

\begin{figure*}[ht]
\vspace{-2em}
\small
\begin{align}
\Upsilon(\alpha) = \frac{e^{-\theta \zeta}\left(\left(e^{-\theta T}\!\!-1\right){H(\alpha)}+1\right)^{-1}}{\left(\left(1-e^{\theta T}\right){H(\alpha)}-1\right)^{-1}+\left( \displaystyle\int_{\frac{\tau N_c}{\alpha P_t}}^{+\infty} \frac{p(\alpha) \eta e^{-h}}{(1-e^{\varpi\theta}(1-p(\alpha)))e^{-\phi(\gamma(\alpha,h))\theta}-p(\alpha)(1-\eta)} \, \mathrm{d}h\right)^{-1}},
\label{t}
\end{align}
\normalsize % 恢复正常字号
where $H(\alpha)=\exp \left(\frac{d(4 \pi)^3 D^{2 \kappa}N_s}{(1-\alpha) P_t \Gamma_t \Gamma_r \sigma^2 \bar{\rho}}\right)$, $ p(\alpha) = \exp (-\frac{\tau N_c}{\alpha P_t})$ and $\gamma(\alpha, h) = \frac{\alpha P_t h}{N_c}$.

\noindent\makebox[\linewidth]{\rule{\paperwidth}{0.4pt}}
\vspace{-2.5em}
\end{figure*}

% \begin{figure*}[ht]
% \vspace{-2em}
% \begin{align}
% \Upsilon(\alpha) &= \frac{e^{-\theta \zeta}\left(\left(e^{-\theta T}\!\!-1\right){H(\alpha)}+1\right)^{-1}}{\left(\left(1\!-\!e^{\theta T}\right){H(\alpha)}-1\right)^{-1}\!+\!\left( \displaystyle\int_{-\!\ln{p(\alpha)}}^{+\infty} \frac{p(\alpha) \eta e^{-h}}{(1-e^{\varpi\theta}(1-p(\alpha)))\exp\!\left(-\phi(\frac{-\tau h}{\ln{p(\alpha)}})\theta\right)-p(\alpha)(1-\eta)} d h\right)^{-1}},
% \label{t}
% \end{align}

% where $H(\alpha)=\exp \left(\frac{d(4 \pi)^3 D^{2 \kappa}N_s}{(1-\alpha) P_t \Gamma_t \Gamma_r \sigma^2 \bar{\rho}}\right)$ and $ p(\alpha) = \exp (-\frac{\tau N_c}{\alpha P_t})$.

% \noindent\makebox[\linewidth]{\rule{\paperwidth}{0.4pt}}
% \vspace{-2.5em}
% \end{figure*}

%\vspace{-0.5em}
\subsection{PAVP Upper Bound Optimization under Power Constraints}

By Eqs. \eqref{i}, \eqref{q} and \eqref{s}, we have the following theorem. 
% Similarly, the statistical PAoI upper bound for the network can be derived using Corollary 1, which is omitted here for brevity.

\textit{Theorem 1:} For a threshold $\zeta$, the statistical PAVP of sensory packets in the ISAC-enabled V2I network can be upper bounded by $\Upsilon(\alpha)$, which is given in Eq. \eqref{t}.

Using Eqs. \eqref{z} and \textit{Theorem 1}, we obtain the following PAVP upper bound minimization problem $\tilde{\mathcal{P}}$:
\begin{equation*}
\begin{aligned}
\tilde{\mathcal{P}}: & \quad \underset{0 \leq \alpha \leq 1}{\min}  && \Upsilon(\alpha) \\
& \,\,\,\,\, \quad \text{s.t.} && \mathbf{M}_{\mathcal{T}^A(n, n+1)}(-\theta) \mathbf{M}_{\mathcal{T}^{\mathcal{S}}(n+1)}(\theta) < 1.
\end{aligned}
\end{equation*}

% $$\tilde{\mathcal{P}}: \min _{0 \leq \alpha \leq 1}  \Upsilon(\alpha).$$

% \begin{equation}
% \begin{array}{ll}
% \tilde{\mathcal{P}}: &\min _{0 \leq \alpha \leq 1}  \Upsilon(\alpha) \\
% \text { s.t. } & \mathbf{M}_{\mathcal{T}^A(n, n+1)}(-\theta) \mathbf{M}_{\mathcal{T}^{\mathcal{S}}(n+1)}(\theta)< 1. 
% \end{array}
% \end{equation}
% Intuitively, from the definition in Eq. (3), PAoI is influenced by two main factors: the total delay experienced by sensory packets during transmission and the interval between the generation of sensory packets. This relationship is encapsulated in our theoretical upper bound framework as presented in Lemma 3, which includes the computation of the MGFs for both packet intervals and service times. In the ISAC-enabled V2I network, the sensing process corresponds to the generation of packets, while the communication process corresponds to their service time. Given the limited resources in such systems, there is an inevitable contention for resources between sensing and communication, leading to a resource allocation challenge.

The convexity of $\Upsilon(\alpha)$ is not immediately apparent due to its complexity. However, through qualitative analysis, we have the following insights: 1) When all power is allocated to communication ($\alpha=1$), the power available for sensing is zero, resulting in an infinitely large interval between sensory packets. Consequently, the numerator of $\Upsilon(\alpha)$ becomes infinite, leading to an infinite $\Upsilon(\alpha)$. This indicates that the PAoI will exceed any practical threshold. 2) Conversely, when all power is allocated to sensing ($\alpha=0$), the power for communication becomes zero, failing to meet the system stability requirement. At the critical value of $\alpha$, where the system stability is just satisfied, the denominator of $\Upsilon(\alpha)$ approaches zero, which again makes $\Upsilon(\alpha)$ infinite. 

Therefore, the trade-off of power between sensing and communication is critical to maintaining the freshness of information, which directly impacts the safety and efficiency of ISAC-enabled V2I networks. By numerically optimizing $\Upsilon(\alpha)$, we obtain optimal power allocation solutions that minimize the statistical PAVP of the sensory packets. These solutions will be presented and validated in the subsequent section, offering valuable insights for the design of such networks.

\vspace{-0.5em}
\section{Numerical Results}

In this section, we perform numerical analysis for the ISAC-enabled V2I network to validate our analysis and evaluate the effects of various parameters. The default parameter settings are listed in TABLE I. 

\begin{table}[t]
\centering
\caption{Default Parameter Settings}
\label{tab:my_label}
\begin{tabular}{cccccc}
\hline
Parameter & Value & Parameter & Value & Parameter & Value \\
\hline
$P_t$ & 10 $\mathrm{W}$ & $\sigma$ & 4 $\mathrm{mm}$ & $\Gamma_t$ & 10 $\mathrm{dBi}$ \\
$\Gamma_r$ & 10 $\mathrm{dBi}$ & $\bar{\rho}$ & 10 $\mathrm{dBsm}$ & $W$ & 25 $\mathrm{kHz}$ \\
$\epsilon$ & 1\textperthousand{} & $N_c$ & -23 $\mathrm{dBm}$ & $T$ & 1 $\mathrm{ms}$ \\
$D$ & 100 $\mathrm{m}$& $\zeta$& 6 $\mathrm{ms}$& $\kappa$& 2\\
$N$& 100& $\varpi$& 0.5 $\mathrm{ms}$& $L$& 100\\
\hline
\end{tabular}
\vspace{-2em}
\end{table}

In Fig. 2, we show the relationship between the SDP and the maximum ranging distance $D$ for various average RCS values $\bar{\rho}$ and sensing SNR thresholds $d$. The markers align closely with the lines, validating the accuracy of our analysis. As $D$ increases, the SDP decreases for both $\bar{\rho}$ values. This trend is expected due to increased path loss with distance. Notably, at 0 dBsm, the SDP decreases more rapidly with distance compared with 10 dBsm. This indicates that objects with higher $\bar{\rho}$ are easier to detect at longer distances. At shorter distances, the difference in SDP between 0 dBsm and 10 dBsm is minimal; however, this gap widens significantly as $D$ increases. 

In Fig. 3, we depict the relationship between the PAVP and the communication bandwidth for various DEPs $\epsilon$. The close alignment of the simulated curves with the theoretical upper bounds confirms the accuracy and reliability of our theoretical framework in predicting system performance. The curves clearly show that increasing the bandwidth effectively reduces PAVP. Additionally, higher $\epsilon$ values result in increased PAVPs across all bandwidths, indicating that systems designed to operate at higher bandwidths are more robust against errors.

In Fig. 4, we show the relationship between the PAVP and the power allocation factor $\alpha$ for various PAoI thresholds. The black circles mark the optimal $\alpha$ values that minimize PAVP. A trade-off between sensing power and communication power is evident, as PAVP initially decreases and then increases with $\alpha$ ranging from 0 to 1, highlighting that an appropriate adjustment of $\alpha$ can significantly reduce PAVP. The PAVP is lower for $\zeta = 4$ compared with $\zeta = 8$, indicating that a higher threshold generally results in better performance. As the threshold decreases, the optimal $\alpha$ point shifts rightward to mitigate the impact of the lower threshold. This trend is well-tracked and reflected by our theoretical bound, underscoring the robustness and accuracy of our analysis.

In Fig. 5, we illustrate the relationship between the PAVP and the threshold value $\zeta$ for different intervals $\varpi$. Theoretical upper bounds closely follow the trends observed in simulated data. It can be seen that for each value of $\varpi$, the PAVP decreases with an increase in $\zeta$. Moreover, as $\varpi$ decreases, the PAVP curves shift downward, indicating that shorter $\varpi$ are associated with lower PAVPs. It is important to note that while a smaller $\varpi$ can improve system performance by allowing faster response to channel changes, $\varpi$ cannot be made arbitrarily small in practical settings. The environmental characteristics of the scenario need to be considered, as smaller $\varpi$ values correspond to more dynamic channel conditions.

\begin{figure*}[htbp]
\vspace{-2.5em}
	\centering
	\begin{minipage}{0.49\linewidth}
		\centering
		\includegraphics[width=0.83\linewidth]{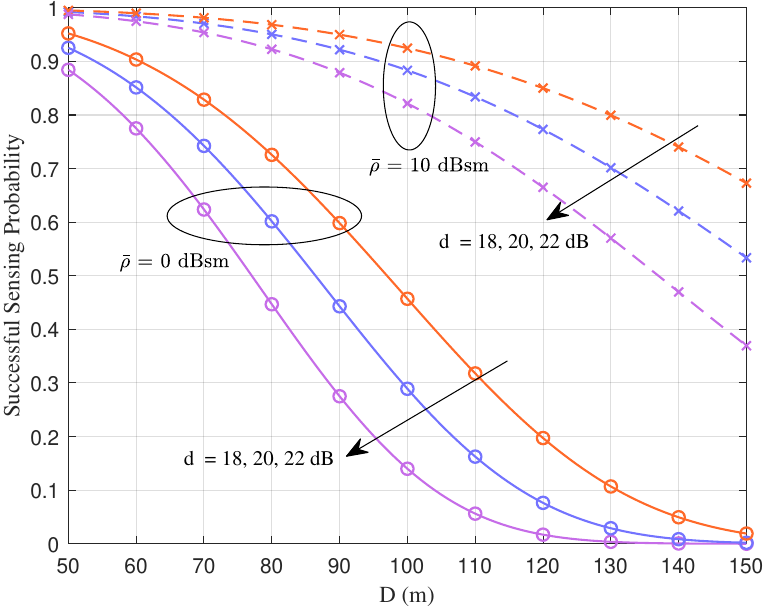}
  \vspace{-0.5em}
  \captionsetup{justification=raggedright,singlelinecheck=false} % 设置标题左对齐
		\caption*{\small Fig. 2: SDP vs. $D$ under various $\bar{\rho}$ and $d$. The lines represent the theoretical values, and the markers represent the simulation values.\\ \phantom{This line is intentionally left blank}}
  \vspace{-0.2em}
		\label{Fig2}%文中引用该图片代号
	\end{minipage}
	\begin{minipage}{0.49\linewidth}
		\centering
		\includegraphics[width=0.83\linewidth]{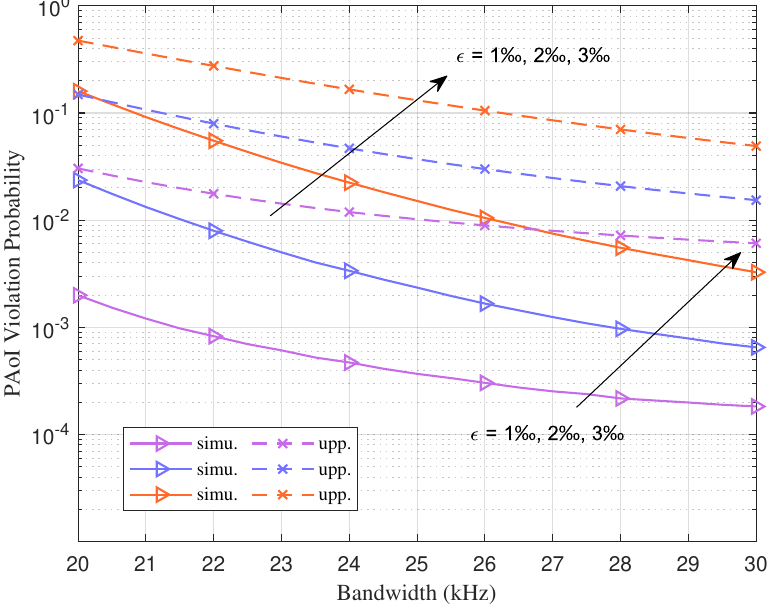}
  \vspace{-0.5em}
  \captionsetup{justification=raggedright,singlelinecheck=false} % 设置标题左对齐
		\caption*{\small \hspace*{0.5em} Fig. 3: PAVP vs. bandwidth under various $\epsilon$. The simulated \hspace*{0.5em} curves, denoted as ``simu.", and the theoretical upper bounds, \hspace*{0.5em} denoted as ``upp.", are used throughout unless otherwise stated.}
  \vspace{-0.2em}
		\label{Fig3}%文中引用该图片代号
	\end{minipage}
 \end{figure*}
 % \\ \phantom{This line is intentionally left blank}
 % The simulated \\ \hspace*{0.5em} curves are denoted as ``simu." and the theoretical upper bounds \\ \hspace*{0.5em} as ``upp.", which are used throughout unless otherwise stated.

 \begin{figure*}[htbp]
 \vspace{-1.3em}
	\begin{minipage}{0.49\linewidth}
		\centering
		\includegraphics[width=0.83\linewidth]{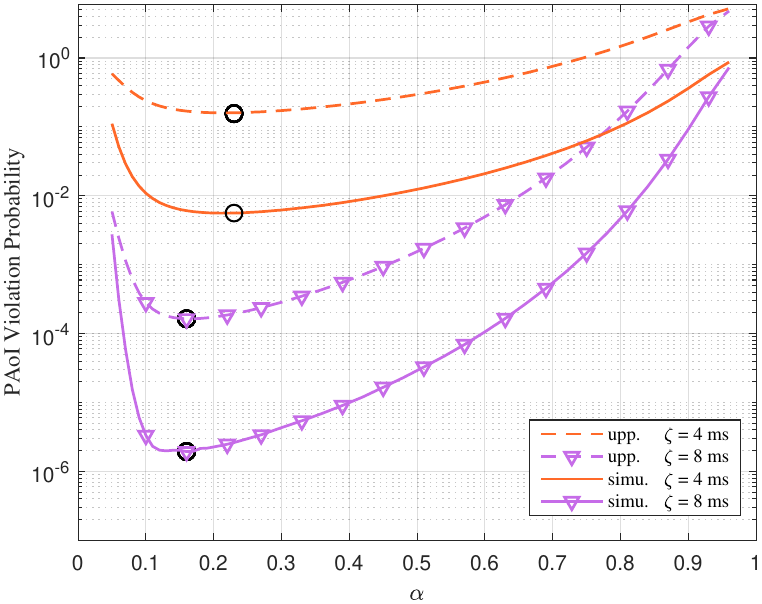}
  \vspace{-0.5em}
  \captionsetup{justification=raggedright,singlelinecheck=false} % 设置标题左对齐
		\caption*{\small Fig. 4: PAVP vs. $\alpha$ under various thresholds $\zeta$.}
		\label{Fig4}%文中引用该图片代号
	\end{minipage}
 \vspace{-1em}
	\begin{minipage}{0.49\linewidth}
		\centering
		\includegraphics[width=0.83\linewidth]{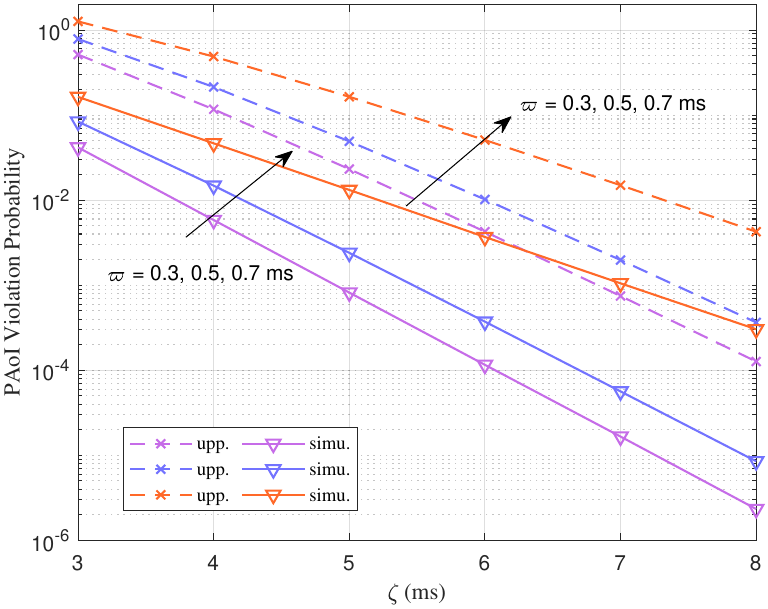}
\vspace{-0.5em}
\captionsetup{justification=raggedright,singlelinecheck=false} % 设置标题左对齐
		\caption*{\small \hspace*{0.5em} Fig. 5: PAVP vs. threshold $\zeta$ under various durations $\varpi$.}
		\label{Fig5}%文中引用该图片代号
	\end{minipage}
\vspace{-0.5em}
\end{figure*}

\vspace{-0.4em}
\section{Conclusion}
This study investigates ISAC-enabled xURLLC V2I networks with short packets under power constraints. Using SNC theory, we develop a theoretical framework that effectively captures network dynamics and derive the PAVP upper bound for the ISAC-enabled V2I network. Optimizing this upper bound leads to a near-optimal power allocation scheme. The alignment between simulations and theoretical upper bounds underscores the effectiveness of the theoretical models used in evaluating system behavior, providing valuable insights for more robust and efficient ISAC-enabled V2I network design.

% Simulations validated that the derived AoI performance upper bound effectively tracks and encompasses the actual AoI performance curve. The proposed power allocation scheme has been validated as near-optimal, effectively minimizing AoI under power constraints. This work provides valuable insights and guidance for the design of ISAC systems, helping to enhance the performance of such networks.

% or

%\clearpage %空白页

% \appendices

% % you can choose not to have a title for an appendix
% % if you want by leaving the argument blank
% \section{}

% \section*{Acknowledgment}


\ifCLASSOPTIONcaptionsoff
  \newpage
\fi



%\bibliographystyle{IEEEtran}
\vspace{-0.3em}

%\bibliographystyle{IEEEtran}

%\bibliography{reference}

\begin{thebibliography}{10}

\bibitem{ZhongEmpoweringV2XNetwork2022}
Y.~Zhong {\em et~al.}, ``Empowering the v2x network by integrated sensing and communications: Background, design, advances, and opportunities,'' {\em IEEE Netw.}, vol.~36, pp.~54--60, July 2022.

\bibitem{LiuIntegratedSensingCommunications2022}
F.~Liu {\em et~al.}, ``Integrated sensing and communications: Toward dual-functional wireless networks for 6g and beyond,'' {\em IEEE J. Sel. Areas Commun.}, vol.~40, pp.~1728--1767, June 2022.

\bibitem{Liu2023Performance}
M.~Liu {\em et~al.}, ``Performance analysis and power allocation for cooperative isac networks,'' {\em IEEE Internet Things J.}, vol.~10, pp.~6336--6351, Apr. 2023.

\bibitem{DuIntegratedSensingCommunications2023}
Z.~Du {\em et~al.}, ``Integrated sensing and communications for v2i networks: Dynamic predictive beamforming for extended vehicle targets,'' {\em IEEE Trans. Wireless Commun.}, vol.~22, pp.~3612--3627, June 2023.

\bibitem{Lopez2023Statistical}
O.~L.~A. L{\'o}pez {\em et~al.}, ``Statistical tools and methodologies for ultrareliable low-latency communication—a tutorial,'' {\em Proc. IEEE}, vol.~111, pp.~1502--1543, Nov. 2023.

\bibitem{Zhao2024Joint}
X.~Zhao and Y.-J.~A. Zhang, ``Joint beamforming and scheduling for integrated sensing and communication systems in urllc: A pomdp approach,'' {\em IEEE Trans. Commun.}, May 2024.

\bibitem{Ding2022Joint}
C.~Ding, C.~Zeng, C.~Chang, J.-B. Wang, and M.~Lin, ``Joint precoding for mimo radar and urllc in isac systems,'' in {\em Proceedings of the 1st ACM MobiCom Workshop on Integrated Sensing and Communications Systems}, (Sydney, NSW, Australia), pp.~12--18, Oct. 2022.

\bibitem{Zhang2021AoI}
X.~Zhang, J.~Wang, and H.~V. Poor, ``Aoi-driven statistical delay and error-rate bounded qos provisioning for murllc over uav-multimedia 6g mobile networks using fbc,'' {\em IEEE J. Sel. Areas Commun.}, vol.~39, pp.~3425--3443, Nov. 2021.

\bibitem{Champati2021Statistical}
J.~P. Champati, H.~Al-Zubaidy, and J.~Gross, ``Statistical guarantee optimization for aoi in single-hop and two-hop fcfs systems with periodic arrivals,'' {\em IEEE Trans. Commun.}, vol.~69, pp.~365--381, Jan. 2021.

\bibitem{Zhong2023Stochastic}
A.~Zhong, Z.~Li, D.~Wu, T.~Tang, and R.~Wang, ``Stochastic peak age of information guarantee for cooperative sensing in internet of everything,'' {\em IEEE Internet Things J.}, vol.~10, pp.~15186--15196, Sept. 2023.

\bibitem{Li2023Effective}
J.~Li {\em et~al.}, ``Effective joint scheduling and power allocation for urllc oriented v2i communications,'' {\em IEEE Trans. Veh. Technol.}, Mar. 2024.

\bibitem{Series2018Systems}
M.~Series, ``Systems characteristics of automotive radars operating in the frequency band 7681 ghz for intelligent transport. systems applications,'' in {\em Recommendation ITU-R}, pp.~2057--1, Jan. 2018.

\end{thebibliography}
% \vspace{-0.2em}

% \input{bare_jrnl.bbl}

% \end{thebibliography}


% \end{IEEEbiography}


% % insert where needed to balance the two columns on the last page with
% %\newpage



%\vfill


\end{document}